# HIGH-FIELD PROPERTIES OF SINGLE-CRYSTALLINE CaVO$_3$

M. H. JUNG,[1,2] I. H. INOUE,[3] H. NAKOTTE,[2] and A. H. LACERDA[1]

[1]*National High Magnetic Field Laboratory, Los Alamos National Laboratory, Los Alamos, New Mexico 87545, USA*

[2]*Department of Physics, New Mexico State University, Las Cruces, New Mexico 88003-8001, USA*

[3]*Electotechnical Laboratory, Tsukuba, Ibaraki 305-8568, Japan*

The magnetic properties of perovskite CaVO$_3$ single crystals have been studied by means of magnetoresistance $\rho(T, H)$ and magnetization $M(H)$ measurements in fields to 18T. At 2 K, the magnetoresistance is positive and a maximum value of $\Delta\rho(18\text{T})/\rho(0) = 16.5\%$ is found for $H//a$. The magnetization exhibits a smooth increase at 2 K, reaching values of $M(18\text{T}) = 0.03, 0.05, 0.17$ $\mu_B$/f.u. for $H//a$, $H//b$, and $H//c$, respectively. This anisotropy found in $M(H)$ is consistent with that observed for $\Delta\rho(H//a) > \Delta\rho(H//b) > \Delta\rho(H//c)$. These results can be interpreted in terms of the field-dependent scattering mechanism of CaVO$_3$.

Corresponding author: Dr. Myung-Hwa Jung
Phone: 505-667-7622, Fax: 505-665-4311
E-mail: mhjung@lanl.gov

Among perovskite-type light-3$d$ oxides, $CaVO_3$ has attracted much attention because of its metallic conductivity (Inoue 1997). This behavior is opposed to the insulating one observed in $LaTiO_3$ (Okada 1993). It was reported on $CaVO_3$ that oxygen defects induce insulating conductivity, while oxygen excess induces metallic conductivity (Fukushima 1994). Another important point of $CaVO_3$ is that the magnetoresistance is positive in the stoichiometric samples and is negative in off-stoichiometric samples (Shirakawa 1995). In addition, there has been no report on the anisotropic properties of $CaVO_3$ so far. Therefore, it is interesting to study the transport and magnetic properties with well-characterized single crystalline samples of $CaVO_3$. The present paper reports on magnetoresistance and magnetization results in fields up to 18 T for stoichiometric $CaVO_3$ single crystals.

Single crystals of $CaVO_3$ were grown by the floating-zone method, details can be found elsewhere (Iga 1992). The experiments under high magnetic fields were carried out utilizing a 20T-superconducting magnet at the National High Magnetic Field Laboratory, Los Alamos Facility in a temperature range between 2 and 300 K. Magnetoresistance was measured by the use of a standard four-probe ac method with Epotek H20E silver epoxy for making electric contacts. The contact resistance was less than 1 Ω. Samples were cut to a rectangular shape with dimensions $1 \times 1 \times 2$ mm$^3$, which allowed accurate measurements of resistivity. The *ac* current was applied along the crystallographic *a* axis and the magnetic field was applied along the *a*, *b*, and *c* axis, respectively. Magnetization measurements were performed with a vibrating sample magnetometer in static fields to 18 T. The data were taken at $T = 2$ K in the configurations for $H//a$, $H//b$, and $H//c$.

In Fig. 1, the electrical resistivity is displayed as a function of temperature on a log-log plot at applied magnetic fields of zero, 10, 18 T. Notice that the resistivity above 100 K is proportional to the equation, $\rho(T) = \rho_o + AT^2$, where the first term ($\rho_o = 0.2$ µΩcm) is attributed to electron-impurity scattering and the second term ($A = 0.003$ µΩcm/K$^2$) is usually attributed to electron-electron scattering. Below 30 K, the electrical resistivity is approximately constant with values of slightly more than 1.2 µΩcm for all three directions, which is the smallest value among previous reported data (Inoue 1997, Fukushima 1994, Shirakawa 1995). This small magnitude of $\rho(T)$ and the large residual resistivity ratio (RRR) ~ 24, indicates that the CaVO$_3$ single crystals examined in the present study contain few defects or impurities. As the temperature is lowered from 100 to 2 K, the magnitude of the resistivity increases when the magnetic field is applied. Thus, we have measured the isothermal magnetoresistance with the applied field along the crystallographic *a*, *b*, and *c* axis.

The normalized magnetoresistance at 2 K is plotted as a function of the applied magnetic field in Fig. 2. It can be noticed that the magnetoresistance is positive but significantly anisotropic $\Delta\rho(H//a) > \Delta\rho(H//b) > \Delta\rho(H//c)$, where $\Delta\rho = \rho(H)–\rho(0)$. Note that $\Delta\rho(H//a)$ and $\Delta\rho(H//b)$ are nearly identical, consistent with the observation in magnetization (see Fig. 3). Our positive magnetoresistance is consistent with the one reported for stoichiometric CaVO$_3$ samples, while negative magnetoresistance is seen in off-stoichiometric samples (Fukushima 1994, Shirakawa 1995). The full-potential linear-augmented-plane-wave (FPLAP) energy-band calculation within local density approximation (LCA) indicates that CaVO$_3$ has three Fermi surfaces, two of which are almost spherical but the third one has an open orbit along the (110) direction (Makino 1998). The presence of the open orbit could account for the anisotropic magnetoresistance as well as the absence of saturation of the positive magnetoresistance.

Figure 3 displays the first measurements of isothermal magnetization at 2 K with the applied field along the *a*, *b*, and *c* axis. The magnetization *M*(*H*) for both *H*//*a* and *H*//*b* is linear to the applied magnetic field, while *M*(*H*) for *H*//*c* shows a tendency toward saturation having a value of 0.2 $\mu_B$/f.u. at 18 T. The relationship between *M*(*H*) and *ρ*(*H*) can be discussed in light of the conduction-electron scattering mechanism of CaVO$_3$. As there exists the open orbit along the (110) direction, we argue that there is larger conduction-electron scattering along the *c* axis. This may result in the smaller increase of *ρ*(*H*//*c*) than that as expected from the magnetic anisotropy in *M*(*H*). In addition, the data for CaVO$_3$ do not follow Kohler's rule, $\Delta\rho(H)/\rho(0) \propto H/\rho(0)$ on a log-log plot indicating a single salient scattering time in the transport (Tinkham 1980).

To conclude, we have studied the magnetic properties of CaVO$_3$ single crystals by the measurements of magnetoresistance and magnetization. At 2 K, the positive magnetoresistance exhibits an anisotropy, $\Delta\rho(H//a) > \Delta\rho(H//b) > \Delta\rho(H//c)$. This is consistent with the results of magnetization measurements, $\Delta M(H//a) < \Delta M(H//b) < \Delta M(H//c)$. Therefore, we conclude that CaVO$_3$ is a system with a strong field-dependent scattering.

The work was supported by a grant from NSF (grant number: DMR-0094241). Work at the National High Magnetic Field Laboratory in Los Alamos was

**Figure captions**

Fig. 1. Temperature dependence of electrical resistivity $\rho(T)$ for $CaVO_3$ between 2 and 300 K at fields 0, 10, and 18 T for $H//a$, $H//b$, and $H//c$ in the logarithmic scale.

Fig. 2. Magnetoresistance $\Delta\rho/\rho(0)$ measured at $T = 2$ K for $CaVO_3$ in the configurations $H//a$, $H//b$, and $H//c$.

Fig. 3. Magnetization $M(H)$ measured at $T = 2$ K for $CaVO_3$ in the configurations $H//a$, $H//b$, and $H//c$.

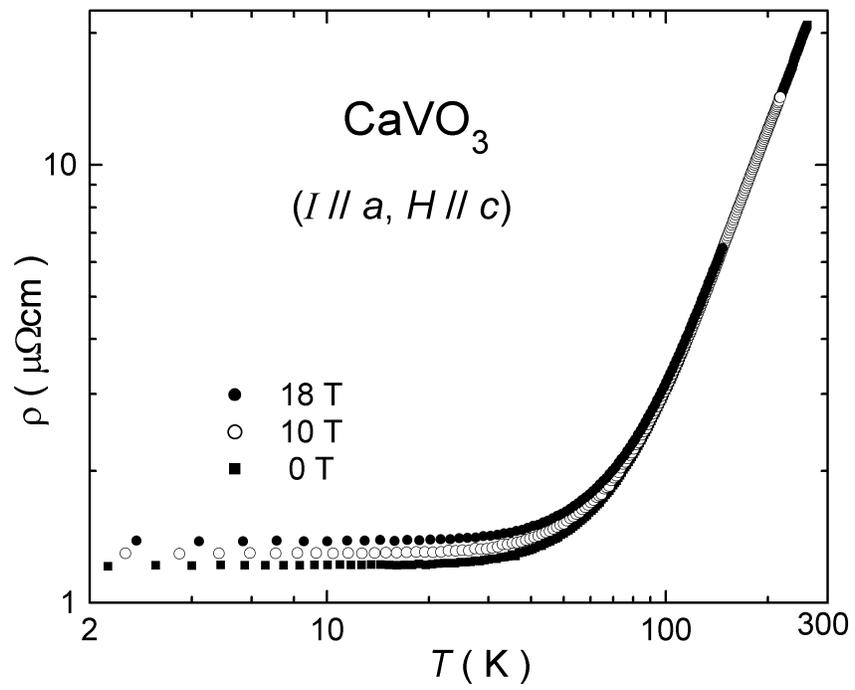

Fig. 1. M. H. Jung et al.

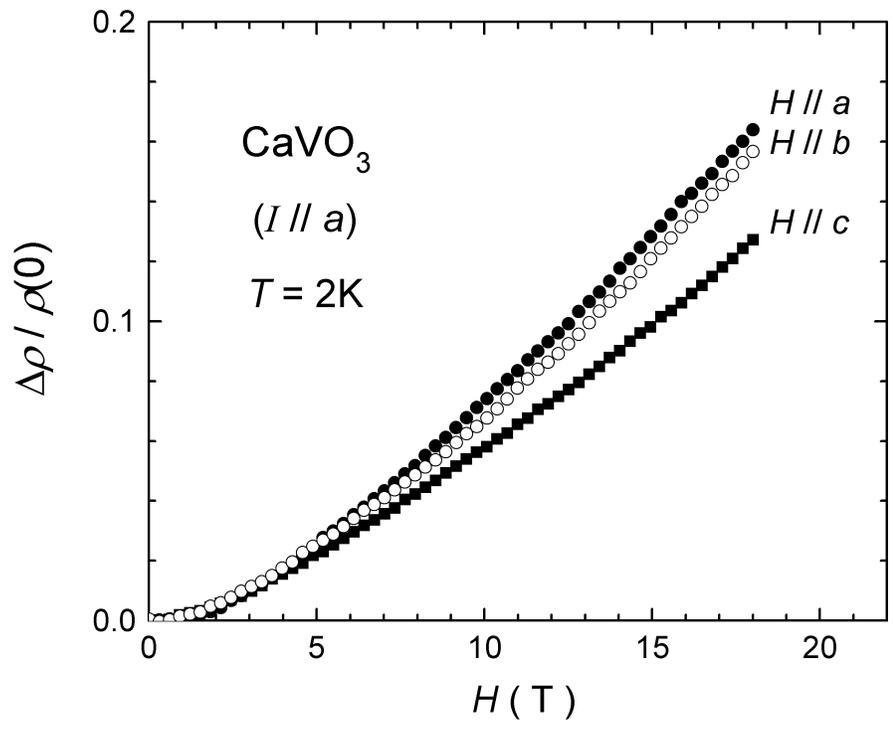

Fig. 2. M. H. Jung et al.

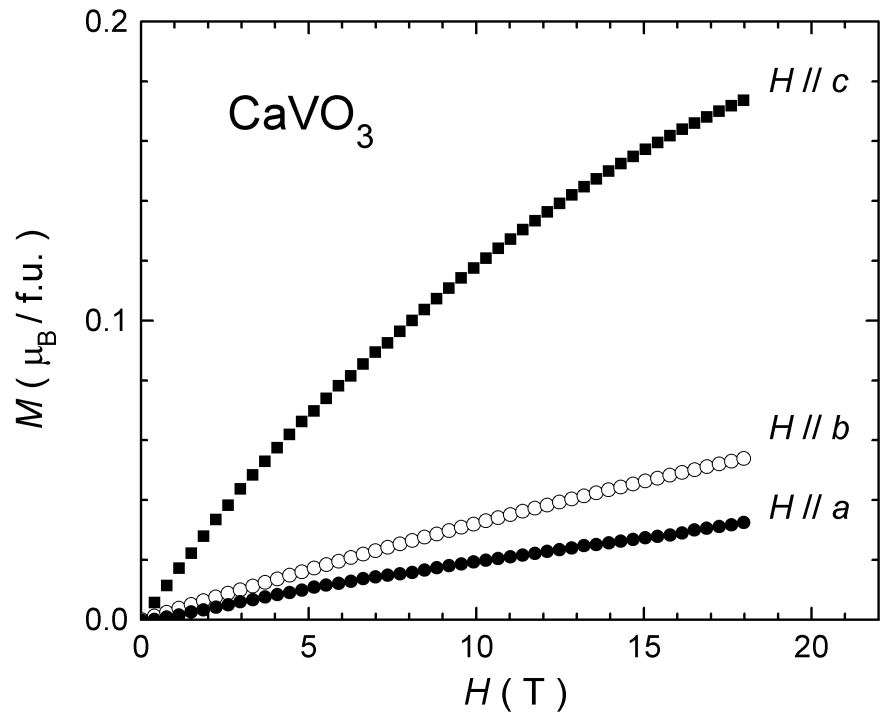

Fig. 3. M. H. Jung et al.